\documentclass[twocolumn,aps,prl,longbibliography,floatfix,10pt]{revtex4-1}

\usepackage[T1]{fontenc}
\usepackage{natbib}
\usepackage{times}
\usepackage{graphicx}
\usepackage{amsfonts}
\usepackage{amsmath,amsthm,amssymb,mathrsfs}
\usepackage{dsfont}
\usepackage{xcolor}
\usepackage{microtype}
\usepackage{bm}
\usepackage{siunitx}
\usepackage[colorlinks=true, allcolors=black, citecolor=blue, linkcolor=blue, urlcolor=blue]{hyperref}
\usepackage[integrals]{wasysym}

\allowdisplaybreaks

\DeclareMathOperator{\Tr}{Tr}

\renewcommand{\rm}[1]{\mathrm{#1}}

\newcommand{\vk}{\mathbf{k}}
\newcommand{\vq}{\mathbf{q}}
\newcommand{\eps}{\epsilon}

\newcommand{\on}{\omega_n}
\newcommand{\On}{\Omega_n}

\newcommand{\F}{\mathrm{F}}
\newcommand{\nF}{n_{\F}}

\newcommand{\su}{\uparrow}
\newcommand{\sd}{\downarrow}

\usepackage{soul}

\usepackage[capitalize]{cleveref}

\begin{document}

\title{Dichroic cavity mode splitting and lifetimes from interactions with a ferromagnetic metal}

\author{Henning G. Hugdal}
\thanks{These two authors contributed equally}
\author{Eirik Jaccheri Høydalsvik}
\thanks{These two authors contributed equally}
\author{Sol H. Jacobsen}
\affiliation{Center for Quantum Spintronics, Department of Physics, NTNU, Norwegian University of Science and Technology, NO-7491 Trondheim, Norway}


\begin{abstract}
We study the effect of ferromagnetic metals (FM) on the circularly polarized modes of an electromagnetic cavity and show that broken time-reversal symmetry leads to a dichroic response of the cavity modes. With one spin-split band, the Zeeman coupling between the FM electrons and cavity modes leads to an anticrossing for mode frequencies comparable to the spin splitting. However, this is only the case for one of the circularly polarized modes, while the other is unaffected by the FM, allowing for the determination of the spin-splitting of the FM using polarization-dependent transmission experiments. Moreover, we show that for two spin-split bands, also the lifetimes of the cavity modes display a polarization-dependent response. The change in photon lifetimes can be understood as a suppression due to level attraction with a continuum of Stoner modes with the same wavevector. The reduced lifetime of modes of only one polarization could potentially be used to engineer and control circularly polarized cavities.
\end{abstract}

\maketitle

In polarization-dependent absorption and transmission experiments, a material is driven by an external circularly polarized light wave. If the transmitted intensity for left- and right-handed photons is different, the material displays circular dichroism (CD) \cite{Ebert1996,Baranov2020CircularCavity-plasmon-polaritons,Capelle1997TheorySuperconductors}. Photons with different chirality are related by time-reversal symmetry (TRS) \cite{Lovesey2014TopologicalSuperconductors}, and CD can therefore be used to study spontaneously broken or induced TRS-breaking systems, e.g. from an external magnetic field. As such, CD has been widely used to gain information about the spin and angular momentum of the electronic bands in ferromagnets \cite{Ebert1996,Schtz1987AbsorptionIron,Funk2005X-rayProperties}.

Resonance experiments in electromagnetic cavities provide an alternative, undriven approach to gaining information about electronic systems \cite{Kittel2005,Wertz1986}, and are used to investigate light-matter coupling \cite{Soykal2010,Huebl2013,Tabuchi2014,Zhang2014,Yuan2017,FriskKockum2019,Forn-Diaz2019,Schlawin2022,ZareRameshti2022}. Cavity photons interact with particles (e.g. electrons, magnons, phonons, excitons) characterizing a material, and may lead to hybridized light-matter states \cite{Carusotto2013,Harder2018,FriskKockum2019,Forn-Diaz2019,ZareRameshti2022,Schlawin2022}. The coupling is inversely proportional to the square root of the cavity volume, and reducing the volume can therefore increase the coupling by several orders of magnitude~\cite{Schlawin2022}. In this way, strong coupling between a cavity and magnetic systems \cite{Huebl2013,Tabuchi2014,Zhang2014,Bialek2021,Boventer2023,Mergenthaler2017} has been achieved experimentally, giving rise to magnon-polariton modes, and may induce attractive electron-electron self-interactions in a normal metal \cite{Schlawin2019,Andolina2024}. Moreover, cavity photons can mediate interactions between spatially separated magnetic systems \cite{Lambert2016,Johansen2018,Nair2022}, ferromagnets and superconductors \cite{Janssonn2020,Janssonn2023}, quantum wells \cite{Sciesiek2020,Fas2021}, and magnets and superconducting qubits \cite{Tabuchi2015,Lachance-Quirion2017ResolvingFerromagnet,Lachance-Quirion2020}, opening possible applications in quantum information \cite{Yuan2022,ZareRameshti2022}.

Light-matter coupling can lead to effective interactions between cavity modes mediated by, e.g., electrons or magnons of a matter-system. This can lead to changes in the cavity resonance frequencies, which can be measured in reflection and transmission experiments~\cite{D.F.Walls1994QuantomOptics,Huebl2013,Tabuchi2014,Zhang2014}. Measuring cavity photons can therefore indirectly probe equilibrium properties of the material placed in the cavity, allowing for the detection of topological phase transitions \cite{Mendez-Cordoba2020}, for example. While electromagnetic cavities utilizing linearly polarized light are ubiquitous, others that excite circularly polarized photons have been developed more recently~\cite{Diaz1974ARadiation,Henderson2008High-frequencySpectroscopy,Watanabe2015CircularlyThesis}. 
We show that these offer new directions for cavity engineering, probing TRS-breaking systems without external drives, and measuring extremely small spin splittings.
\begin{figure}[b]
    \includegraphics{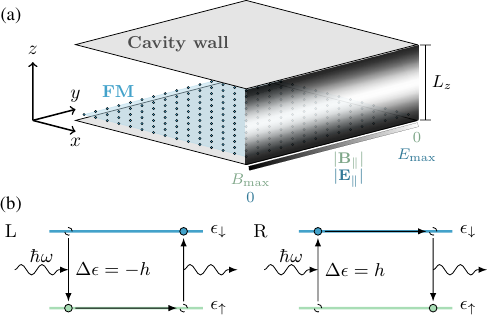}
    \caption{\label{fig:system_diagram} (a) Electromagnetic cavity containing a thin FM. Variation of in-plane magnitudes for the magnetic and electric field is illustrated for the $n_z=1$ modes. (b)
    The change in energy is opposite when scattering left- (L) and right- (R) circularly polarized photons on spin-up ($\su$) and -down ($\sd$) electrons, allowing only one polarization to interact resonantly.}
\end{figure}%

In particular, we study circular dichroism associated with cavities interacting with spin-split metallic systems, such as a ferromagnetic metal (FM), [see \cref{fig:system_diagram}(a)].
Using a single-band model, we show that cavity modes with energies close to the spin splitting couple strongly to electrons for one polarization. This gives an anticrossing in the spectrum, while the oppositely polarized modes are unaffected by the electronic system. Hence, the resonance frequencies are split in a similar way as for magnons \cite{Proskurin2019} and excitons \cite{Zhumagulov2023} in spin-split systems, and as was recently observed experimentally \cite{Chen2022a}.
By introducing a second spin-split band, we show that, despite the same underlying dichroic mechanism [\cref{fig:system_diagram}(b)], a difference in effective masses for the bands leads to a range of allowed transition frequencies for zero momentum transfer. In addition to renormalized resonance peaks, this gives polarization-dependent suppression of the cavity modes over a wide range of system parameters, not possible when coupling to magnons or excitons. Hence, we suggest that interactions between cavity modes and spin-split metals can be used to uniquely control the difference in resonance frequency and lifetime between left- and right-handed cavity modes. This paves the way for cavity devices with large, magnetically tunable CD signals, controlling spin signals \cite{Yuan2017,Liu2022b,Matsubara2022} and magnon entanglement \cite{Yuan2020a} for spintronics and quantum information applications, frequency-dependent linear-to-circular polarization converters \cite{Wang2017a,Valagiannopoulos2017,Sarsen2019}, and non-invasive cavity engineering \cite{Hubener2021}.

\paragraph{Theory.}
We consider a FM placed in a rectangular electromagnetic cavity with dimensions $L_z \ll L_x, L_y$ [\cref{fig:system_diagram}(a)]. Due to the variation of the electric and magnetic field components in the cavity, the vertical placement of the FM determines the relative strength of the coupling channels between cavity and FM-electrons. We write the cavity theory in a circularly polarized basis, and consider how the coupling to the FM compares for spin-split single and two-band models.

The system is described by the imaginary time action $S = S_\mathrm{p} + S_\mathrm{el} + S_\mathrm{int}$, where the first two terms represent the non-interacting photons and electrons, parameterized by complex fields $a_{ql}^{(\dagger)}$ and Grassmann fields $c_{kb}^{(\dagger)}$, respectively \cite{Altland2010,Tsvelik2003}. Here, $l \in \{\text{L},\text{R}\}$ denotes the left- and right-handed circular polarization, and we have included the frequency-dependency of the cavity fields in the four-vector $q = \left(\Omega_n, \vq\right)$, with bosonic Matsubara frequency $\Omega_n = 2n\pi/\beta$ \cite{Altland2010} and wavevector $\vq$. Here, $n\in \mathbb{Z}$ and the inverse temperature $\beta = \hbar/k_\rm{B} T$ , with Planck constant $\hbar$, Boltzmann constant $k_\rm{B}$ and system
temperature $T$. For the electron fields, the index $b=(\sigma,\nu)$ classifies spin up(down) $\sigma=\pm 1$ and band number $\nu$, and $k = (\on, \vk)$  with fermionic Matsubara frequency $\omega_n=(2n+1)\pi/\beta$, and wavevector $\vk$.

A general interaction between electrons and cavity photons to first order in bosonic fields is
\begin{align}
    S_\mathrm{int} = \frac{1}{\sqrt{\beta}}\sum_{kk'}\sum_{lbb'} g_{\vk-\vk'}^{lbb'} c_{kb}^\dagger c_{k'b'}(a_{k-k'l} + a_{k'-k\bar{l}}^\dagger),
\end{align}
where $\bar{L}(\bar{R}) = R(L)$, with time-reversal symmetric interaction matrix, $(g_{\vq}^{lb'b})^\dagger = g_{-\vq}^{\bar{l}bb'}$. For simplicity we assume the interaction does not depend on the electron momentum $\vk$. Moreover, since we are interested in the dichroic response of photons to the presence of an FM, we exclude interactions with $b=b'$, which couple equally to L- and R-photons in the dipole limit, $\vq \to 0$. The full system action is then \cite{Kakazu1994,Janssonn2023}
\begin{align}
    S = \sum_{q l} (-i\hbar \Omega_n + \hbar \omega_{\vq}) a_{ql}^{\dagger}a_{ql} - \sum_{k k'bb'} c_{kb}^{\dagger} \mathcal{G}^{-1}_{kk';bb'} c_{k'b'},
    \label{system action nambu}
\end{align}
with photon dispersion $\omega_{\vq} = c|\vq|$, speed of light $c$, and FM electron Green's function
\begin{align}
    \mathcal{G}^{-1}_{kk';bb'} = (\mathcal{G}_0^{-1} + \chi)_{kk';bb'} = G_{kb}^{-1}\delta_{kk'}\delta_{bb'} + \chi_{kk';bb'},
\end{align}
where
\begin{align}
    G^{-1}_{k b} ={}& i\hbar\on - \epsilon_{\vk b},\\
    \chi_{kk';bb'} ={}& -\frac{1}{\sqrt{\beta}}\sum_l g^{lbb'}_{\vk-\vk'} (a_{k-k'l} + a_{k'-k\bar{l}}^\dagger).
\end{align}
Electron single-particle energies $\eps_{\vk b}$ will be specified for the one- and two-band models.

The effective photon theory can be analyzed by integrating out the electronic bands. Performing the fermionic path integral \cite{Altland2010}, we get the effective action
\begin{align}
    S_\mathrm{eff} = - \hbar \Tr \ln(-\beta \mathcal{G}^{-1}/\hbar),
\end{align}
where the trace is to be taken over all electron degrees of freedom. Using the identity $\mathcal{G}^{-1} = \mathcal{G}_0^{-1}\left(1 + \mathcal{G}_0 \chi \right)$, one can expand the logarithm in powers of $\mathcal{G}_0\chi$ \cite{Altland2010}. To second order in the electron-photon interaction strength, we get the photon-photon interaction \cite{Supp} 
\begin{align}
    \delta S_\mathrm{p} = \frac{\hbar}{2\beta}\!\!\sum_{qkll'bb'} \!\! \frac{g_{\vq}^{lbb'}g_{-\vq}^{l'b'b}(a_{ql} + a_{-q\bar{l}}^\dagger) (a_{-ql'} + a_{q\bar{l}'}^\dagger)}{(i\hbar\on + i\hbar\On - \eps_{\vk+\vq b})(i\hbar\on-\eps_{\vk b'})}. \label{eq:delta_S}
\end{align}
This leads to various coupling terms between the cavity modes, with the exact form depending on the matrix elements $g_{\vq}^{lbb'}$.

\paragraph{Spin-split single-band model.}
An FM is often modeled as a free-electron gas, with exchange splitting $h$ giving the energy difference between spin directions $\sigma$ \cite{Nolting2009},
\begin{align}
    \epsilon_{\vk \sigma} = \frac{\hbar^2 \vk^2}{2 m}-\mu - \sigma \frac{h}{2} \equiv \epsilon_{\vk} - \sigma \frac{h}{2}, \label{eq:eps_one_band}
\end{align}
with chemical potential $\mu$ and electron mass $m$. When the FM is placed at $z\approx 0$, the relevant electron-phonon interaction is due to the Zeeman coupling,
\begin{align}
    g_\vq^{l\sigma\sigma'} = \tilde{g}_\vq \delta_{lL}\delta_{\sigma\su}\delta_{\sigma'\sd} + \tilde{g}_{-\vq}^* \delta_{lR}\delta_{\sigma \sd}\delta_{\sigma'\su}, \label{eq:g_one_band}
\end{align}
with coupling strength $\tilde{g}_\vq$.
Inserting into \cref{eq:delta_S}, the effective photon theory and inverse Green's function become
\begin{align}
    {S}_\mathrm{p}^\mathrm{eff} ={}& -\sum_{q} \begin{pmatrix}
        a_{q\textrm{L}}^\dagger & a_{-q\textrm{R}}
    \end{pmatrix}\mathcal{D}_q^{-1}\begin{pmatrix}
        a_{q\textrm{L}} \\ a_{-q\textrm{R}}^\dagger
    \end{pmatrix},\label{eq:S_eff_matrix} \\
    \mathcal{D}_q^{-1} ={}& \begin{bmatrix}
        i\hbar\On - \hbar \omega_\vq - \Pi_q & -\Pi_q \\
        -\Pi_q & -i\hbar\On - \hbar\omega_\vq - \Pi_q
    \end{bmatrix},\label{eq:D_q_inv}
\end{align}
with self-energy $\Pi_q$, to be determined. Note that R-fields enter with negative four-vector argument here, which for an inversion symmetric system is equivalent to a negative frequency argument.

${S}_\mathrm{p}^\mathrm{eff}$ comprises two types of self-energy terms. The first describes processes that shift the resonance frequency of the renormalized photons $a_{q L}^{\dagger} a_{q L}$ and $a_{-q R}^{\dagger}a_{-q R}$. The second, off-diagonal terms in $D_q^{-1}$, create/annihilate photon pairs with opposite handedness and momentum, $ a_{-q R}^{\dagger}a_{q L}^{\dagger}$ and $a_{q L}a_{-q R}$. This hybridizes the photons, leading to new cavity states with mixed polarization, with eigenstates found by diagonalization. To probe dichroism, we derive the effective theory for the L- and R-photons, which are the states probed in transmission experiments involving only one photon polarization.

Changes to the resonance frequencies of the circularly polarized states can be found from the photon Green's functions. Inverting the matrix in \cref{eq:D_q_inv}, we find the renormalized Green's functions for the L- and R-modes \cite{Altland2010},
\begin{align}
    D_q^\mathrm{L/R} = \frac{i\hbar\On + \hbar\omega_\vq + \Pi_{\pm q}}{(i\hbar\On)^2 - (\hbar\omega_\vq)^2 - 2\hbar\omega_\vq\Pi_{\pm q}}.\label{eq:D_LR}
\end{align}
If $\Pi_q$ is even in $q$, $D_q^\mathrm{L(R)}$ become identical. A necessary condition for CD is therefore that the photon self-energy $\Pi_q \neq \Pi_{-q}$. In the dipole approximation, $\vq\to0$, this simplifies to $\Pi_q$ not being even in frequency.

Using \cref{eq:eps_one_band,eq:g_one_band}, we find
\begin{align}
    \Pi_q ={}& \frac{\hbar}{\beta} |\tilde{g}_{\vq}|^2 \sum_{k} 
    G_{k+q\su} G_{k\sd} 
    = -  \frac{|\tilde{g}_\vq|^2\delta N}{i \hbar \Omega_{n} + h}.
\label{eq:Piq_FM}
\end{align}
In the last equality we inserted the bare electron Green's functions $G_{k \sigma}$, performed the Matsubara summation \cite{Altland2010}, and used the dipole approximation in the electron dispersion relations. We also defined the difference in occupation between spin-up and spin-down bands, $\delta N =  \sum_{\vk} [ n_{\mathrm{F}}(\epsilon_{\vk\su}) - n_{\mathrm{F}}(\epsilon_{\vk\sd})],$ with Fermi-Dirac distribution $n_{\mathrm{F}}(\epsilon) = [\exp(\beta\epsilon/\hbar) + 1]^{-1}$. When both bands are partly filled, $\delta N \propto h$ to leading order. 
Inserting \cref{eq:Piq_FM} into \cref{eq:D_LR},
and analytically continuing to real frequencies, $i\On \to \omega + i\delta$ with $\delta = 0^+$, yields photon propagators $D^l_q$ expressed in real frequencies. The resonances of the cavity photons are given by the roots of the denominator of the propagators, a cubic equation in $\omega$. However, since the resonances are at $\omega = \omega_\vq \equiv \omega_0$ when the light-matter coupling is zero, we neglect negative frequency solutions by letting $\omega + \omega_\vq \to 2\omega_\vq$. The root equation reduces to second order in frequency, with solutions
\begin{align}
    \omega_\pm =  \frac{\omega_{\vq}}{2} - l \frac{\omega_h}{2} \pm \frac{1}{2} \sqrt{\left(\omega_{\vq} + l\omega_h\right)^2 - 4l\kappa\omega_h},
    \label{eq:omega_FM}
\end{align}
where $\omega_h \equiv h/\hbar$ is the spin-splitting frequency, and $\kappa \equiv |\tilde{g}_\vq|^2 \delta N/h\hbar$ is the effective photon-photon coupling. 

\begin{figure}
\includegraphics[width=\columnwidth]{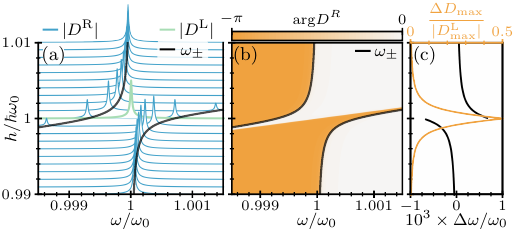}
\caption{\label{fig:FM_res_vs_h} (a) Absolute value and (b) phase of the propagator \eqref{eq:D_LR} for R-photons (blue), with $\delta/\omega_0 = 10^{-5}$ and $\kappa/\omega_0 = \num{5e-7}$, compared with the analytic solution \eqref{eq:omega_FM} (black).  For spin-splitting close to $\hbar\omega_0$ there is an anticrossing, shifting the resonance frequencies away from $\omega_0$.  L-photons (green) are unaffected by the FM. (c) Difference between dominant L- and R-photon resonance frequencies (black) and peak heights (orange).}
\end{figure}

\cref{eq:omega_FM} shows that cavity photons interacting with a spin-split electron band have two separate resonances for left-handed ($l = 1$) and right-handed ($l = -1$) photons, and hence a polarization-dependent transmission spectrum. Therefore, the system displays CD for sufficiently strong coupling $\kappa$ when $\omega_\vq \sim \omega_h$. This is evident in \cref{fig:FM_res_vs_h}(a), showing the resonance frequencies [\cref{eq:omega_FM}] and exact propagators [\cref{eq:D_LR}]. The analytical and numerical results are in excellent agreement, showing an anticrossing for R-photons when $h\sim \hbar\omega_0$, and none for L-photons, leading to a finite peak splitting and different peak intensities [\cref{fig:FM_res_vs_h}(c)]. The anticrossing can be understood as a hybridization between cavity modes and a zero-momentum Stoner boson \cite{Nolting2009,Zakeri2014}, and is therefore qualitatively similar to coupling to other bosonic modes, i.e., magnons \cite{Proskurin2019} and excitons \cite{Zhumagulov2023} in TRS-broken systems.

The difference between the L- and R-photon spectra is explained by the on-shell scattering process illustrated in \cref{fig:system_diagram}(b): An R-photon can flip electron spin from up to down. Hence, for $\epsilon_\sd - \epsilon_\su = h > 0$, these can interact resonantly with the FM when $\hbar\omega \sim h$, leading to an energy shift. However, L-photons flip electron spin from down to up, and the resulting change in electron and photon energy have opposite sign. Hence, the resonance condition can never be fulfilled, leaving L-photons unaffected by the presence of the FM. The strong interaction and avoided crossing is due to the fact that the energy difference between the spin bands is constant, leading to a simple pole in the photon self-energy, \cref{eq:Piq_FM}.

\paragraph{Two spin-split bands.}
The only light-matter interaction relevant for dichroism with a single band is the magnetic dipole coupling. Since electric dipole coupling is several orders of magnitude stronger \cite{ZareRameshti2022,Janssonn2023}, additional bands may lead to stronger dichroism. Different effective masses of multiple bands yield momentum-dependent energy gaps and more complicated photon self-energies. In addition to the spin-split band $\eps_{\vk\sigma}$ $(\nu = 1)$ in \cref{eq:eps_one_band}, consider a second band,
\begin{align}
    \xi_{\vk\sigma} ={} \frac{\hbar^2\vk^2}{2m}\gamma - \mu - \frac{H}{2}\sigma + \Delta,~(\nu = 2),
\end{align}
where $\gamma$ determines the relative effective mass, and $H$ and $\Delta$ are the exchange splitting and energy shift of the second band. As an example, we assume that L-photons scatter spin-down electrons in one band ($\nu$) to spin-up states in the other band ($\bar{\nu}$), while R-photons scatter electrons from spin-up to spin-down states in different bands \cite{Note1}
\begin{align}
g_{\vq}^{l\sigma\sigma'\nu\nu'} = \delta_{\bar{\sigma}'\sigma}\delta_{\bar{\nu}'\nu}\left[\tilde{g}_\vq \delta_{lL}\delta_{\sigma'\sd} + \tilde{g}_{-\vq}^* \delta_{lR}\delta_{\sigma'\su}\right].
\end{align}
Here, we have neglected intraband spin-conserving scattering, since this would affect both photon polarizations in the same way. Inserting this interaction matrix into \cref{eq:delta_S} and performing the Matsubara frequency summation, we get the self-energy
\begin{align}
    \Pi_q ={}& -\sum_{\vk\sigma} |\tilde{g}_\vq|^2 \sigma \frac{\nF(\eps_{\vk+\vq\sigma}) - \nF(\xi_{\vk\bar{\sigma}})}{i\hbar\On - \sigma(\eps_{\vk+\vq\sigma} - \xi_{\vk\bar{\sigma}})}.
\end{align}
Performing the momentum sums at $T=0$, after analytically continuing to real photon frequencies $\omega$ \cite{Tsvelik2003}, we find the self-energy \cite{Supp}
\begin{align}
        \Pi_{lq}\! ={}& \hbar{\kappa} \!\sum_{\sigma\nu} \!
    \frac{(-1)^\nu}{{1-\gamma}} \!\sqrt{\frac{\smash[b]{\eps_{\nu\sigma}}}{\smash[b]{\mu}}}\left[1 - \sqrt{\frac{\smash[b]{\Theta_\sigma}}{\smash[b]{\eps_{\nu\sigma}}}}\arctan\!\sqrt{\frac{\smash[b]{\eps_{\nu\sigma}}}{\smash[b]{\Theta_\sigma}}}\right]\!. \label{eq:Pi_q_two_band}
\end{align}
Here we have used the dipole approximation in the electron dispersion relations, and defined the energies $\eps_{1\sigma} = \mu + \frac{h\sigma}{2}$, $\eps_{2\sigma} = (\mu - \Delta - \frac{H\sigma}{2})/\gamma$, and
\begin{align}
\Theta_{\sigma} ={}& \frac{\sigma}{\gamma-1}\left(l\hbar\omega + il\delta + \frac{h + H}{2} + \sigma\Delta\right).
\end{align}
The coupling strength $\kappa \equiv \sqrt{2\mu m^3}|\tilde{g}_\vq|^2V_\mathrm{el}/\pi^2\hbar^4$, where $V_\mathrm{el}$ is the FM volume.

\begin{figure}
    \includegraphics[width=\columnwidth]{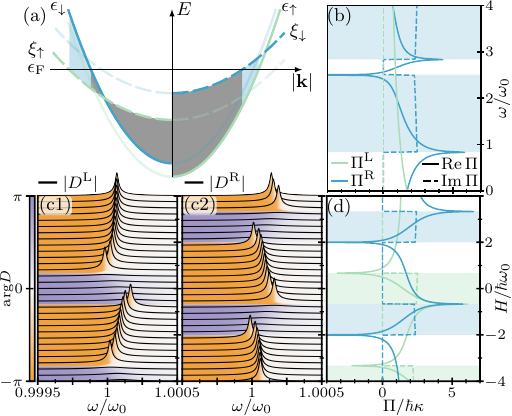}
    \caption{\label{fig:Pi_diagram_two_band} (a) Band dispersion and (b) resulting self-energy $\Pi(\omega)$ for L- and R-photons when 
    $\mu/\omega_0 = 15$, $\Delta/\omega_0 = 9$, $h=2\hbar\omega_0$, $H=3\hbar\omega_0$, $\gamma = 1/3$, and $\delta/\omega_0=10^{-5}$.
    Blue (gray) shading show regions of allowed (suppressed) scattering for R-photons, while there are no allowed regions for L-photons for these parameters. (c) Photon resonances (black lines) and phase (purple/orange shading, see colorbar) vs. spin-splitting $H$, and (d) corresponding self-energies evaluated at the bare cavity frequency $\omega_0$, with $\kappa/\omega_0 = \num{5e-5}$ and parameters as above. The $2\pi$ phase jumps in (c) coincide with the onset of regions where L or R modes are suppressed, indicated by green or blue shading in (d).}
\end{figure}

The expression in square brackets in \cref{eq:Pi_q_two_band} is complex for $-\eps_{\nu\sigma} < \Theta_{\sigma} < 0$, with diverging real part as $\Theta_{\sigma} \to -\eps_{\nu\sigma}$. To understand the physical origin of this behavior, we solve for $\omega$ in the two cases $\operatorname{Re} \Theta_\sigma = (0,-\eps_{\nu\sigma})$. In the former, the frequencies $\hbar\Omega_{0\sigma} = -l\sigma\left(\frac{h + H}{2}\sigma + \Delta\right) $, corresponding to energy difference $\eps_{\vk\sigma} - \xi_{\vk\bar{\sigma}}$ for $\vk = 0$. In the latter,
\begin{align}
    \hbar\Omega_{\nu\sigma} ={}& l\sigma \left[(1 - \gamma)\mu - \Delta - \frac{\gamma h + H}{2}\sigma\right]\gamma^{1-\nu}, \label{eq:Omega_nu_sigma}
\end{align}
corresponding to the energy difference evaluated at the Fermi level of the various bands. This reflects the fact that a photon can resonantly scatter an electron from one band to another if the photon frequency is within the range of energy differences between the initial and final band, as illustrated by the shaded regions between bands in \cref{fig:Pi_diagram_two_band}(a). Moreover, a scattering process is not allowed if the final state is below the Fermi level $\epsilon_\mathrm{F}$ [gray areas in \cref{fig:Pi_diagram_two_band}(a)], restricting scattering to positive frequencies between $\Omega_{1\sigma}$ and $\Omega_{2\sigma}$. For R-modes this corresponds to the blue-shaded areas, while there are no allowed resonant scattering processes for L-modes.

The variation of the photon self-energy with photon frequency is shown in \cref{fig:Pi_diagram_two_band}(b), where the blue shaded areas correspond to the range of allowed frequencies shown in \cref{fig:Pi_diagram_two_band}(a). The self-energy for R-photons has diverging real part at the edges of the ranges of allowed frequencies, and finite imaginary part within the intervals due to the different effective masses, thus reducing the lifetime of the cavity modes. For L-photons, with no range of allowed frequencies for these specific parameters, the self-energy is neither complex nor diverging. Generally, both L- and R-photons can have ranges of allowed frequencies. For identical effective masses ($\gamma = 1$), we have $\Omega_{\nu\sigma} = \Omega_{0\sigma}$, and the self-energy has simple poles at $\omega = \Omega_{0\sigma}$, with effective coupling strength determined by the differences in particle numbers between the bands. Therefore, this special case can yield similar results to the one-band model.

For certain frequency ranges, the new self-energies, affected by a difference in effective masses, has interesting consequences for the response of the cavity photons. For $\gamma = 1$, the simple pole gives a very narrow non-trivial frequency range, set by the effective coupling $\kappa$. For shifts of the photon resonances, which depend on the real part of the self-energy, this is also the case when $\gamma \neq 1$. However, the multiple relevant frequencies $\Omega_{\nu\sigma}$ [\cref{eq:Omega_nu_sigma}] give more possibilities for achieving a good match between the cavity and spin-split material, as seen in \cref{fig:Pi_diagram_two_band}(c). Moreover, the imaginary part of the self-energy [\cref{fig:Pi_diagram_two_band}(d)] is potentially finite within a \emph{range} of frequencies, and could lead to suppressed lifetimes for a large number of cavity modes or specific modes over a wide range of system parameters. This is illustrated in \cref{fig:Pi_diagram_two_band}(c), where either the L- \emph{or} R-modes are suppressed for a range of values of spin-splitting $H$, evidenced by the absence of resonance peaks coinciding with the green (blue) shaded regions in \cref{fig:Pi_diagram_two_band}(d) indicating finite imaginary self-energies for L (R) modes. This enhanced decay is reminiscent of both the Purcell effect \cite{Purcell1946,D.F.Walls1994QuantomOptics}, previously found when coupling cavity photons to lossy magnons \cite{Zhang2014}, and the level-attraction observed due to dissipative magnon-photon coupling \cite{Harder2018a,Yang2019b,Bhoi2019}. These two effects can be distinguished by the phase of the propagator \cite{Harder2018a,Proskurin2019,Yu2019}, and the $2\pi$ phase jumps seen in \cref{fig:Pi_diagram_two_band}(c) indicate that we have level attraction between the cavity modes and Stoner excitations in the FM \cite{Supp}. Note that imaginary self-energies are independent of $\delta$, and the frequency range of suppressed cavity modes can far exceed the coupling constant, making this distinctly different from level attraction observed due to interactions with magnons. Moreover, the parameters for the electronic system can differ from the cavity frequency by orders of magnitude and still give $\Omega_{\nu\sigma} \sim \omega_\vq$ for sufficiently different effective masses. These bands give rise to a continuous range of Stoner excitation frequencies at zero momentum transfer \cite{Kirschner1984,Hopster1984,Abraham1989,Gokhale1994}, and shows the qualitative difference in coupling to a continuum of, or discrete bosonic modes. Our results therefore illustrate the unique possibilities of engineering the cavity modes using spin-split metals, allowing tuning of the resonance frequencies and lifetimes, with controlled suppression of one polarization in certain frequency ranges, with applications in linear-to-circular polarization converters, cut-off filters, and engineering of quantum materials \cite{Hubener2021}.

\paragraph{Concluding remarks.}
We have shown a dichroic effect on  cavity photons when coupling to a metal with spin-split electron bands. The resonance peaks of modes with opposite polarizations can be shifted independently, and controlling and probing the circularly polarized cavity modes separately should enable accurate estimation of the spin splitting without subjecting the FM itself to external drives. For relatively weak spin splitting $h\sim\SI{1}{m eV}$~\cite{Jacobsen2015CriticalCoupling,Bode1999,Xiong2022,Elmers2023}, the resonance condition in the one-band case gives $\omega_0 / 2\pi \sim \SI{250}{GHz}$ and mode splitting $\left(\omega - \omega_0\right) / 2\pi \approx \SI{250}{MHz}$, achievable and observable with current techniques \cite{Schlawin2022,Huebl2013}.

When the effective masses of interacting bands differ, the polarization-dependent suppression of cavity mode lifetimes could allow separate tuning of resonance frequencies and lifetimes of cavity modes with different polarizations. One has inherent freedom in tuning the relevant cavity frequencies, and \emph{in-situ} parameter-control may also be possible, for example with an external voltage controlling the effective spin-orbit coupling, which dictates the relative effective masses of the bands \cite{Manchon2015}.

Our results apply not only to intrinsically spin-split materials, but also to metals with externally broken TRS. Hence, applying a static magnetic field could be used to tune either an FM or a non-magnetic metal to achieve a dichroic response from the cavity modes, allowing direct control over the shifts in resonance frequency and an on/off switch for modes of one polarization. This could potentially control, e.g., chiral magnons \cite{Liu2022b} and spin-polarized photocurrents \cite{Matsubara2022}, as well as selective coupling to specific magnon modes \cite{Yuan2017} and magnon-magnon entanglement \cite{Yuan2020a} in individual magnonic systems, and coupling between different subsystems in cavity spintronics devices. Moreover, such cavity control is essential for equilibrium engineering of cavity quantum materials \cite{Hubener2021}.

We have focused on the overall frequency shift ($\vq = 0$) of the cavity modes, requiring TRS-breaking of the momentum-averaged electronic bands and thus a finite magnetization. However, other magnetic systems with zero magnetization could lead to a dichroic $\vq$-dependence of the photon dispersion relations. For instance, the $d$-wave like spin-splitting in altermagnets \cite{Smejkal2022,Smejkal2022a} could lead to polarization- and direction-dependent photon dispersions.

A recent experiment on metal--insulator--metal interference microcavities with ferrimagnetic upper layers found a magnetically tunable frequency splitting, evident from large magnetic circular dichroism (MCD) signals \cite{Chen2022a}, in agreement with our results. They also show strong dependence on the composition of the ferrimagnetic material, which could be explained by the non-trivial dependence of peak positions and heights on the material parameters. The authors speculate that the difference in signal peaks for L- and R-photons is due to differences in absorption. Our results indicate that the difference in absorption could be due both to the splitting of resonance peaks for one polarization, \textit{or} a decrease in photon lifetimes [see \cref{fig:FM_res_vs_h,fig:Pi_diagram_two_band}]. Finally, we speculate that one may achieve even higher MCD signals in microcavities by placing a ferri- or ferromagnet in the middle of a microcavity, between two dielectric layers. This could lead to stronger coupling overall, as well as coupling between bands with different effective masses, allowing for separate magnetic control of polarized photon lifetimes. The resulting reduction (increase) in transmission (reflection) for one polarization would give a large increase in MCD signal compared to simply splitting the resonances.

\begin{acknowledgments}
\paragraph{Acknowledgments.}
H.G.H. thanks M. Amundsen, B. Brekke, S. D. Lundemo and F. Schlawin for useful discussions. We acknowledge funding via the ``Outstanding Academic Fellows'' programme at NTNU and the Research Council of Norway Grant Nos. 302315 and 262633.
\end{acknowledgments}


%

\end{document}